# Extreme energy density confined inside a transparent crystal:
# Solid - plasma - solid transformations, status and perspectives

**E.G. Gamaly[1], S. Juodkazis[2], A.V. Rode[1]**

[1]*Laser Physics Centre, Research School of Physics and Engineering, The Australian National University, Canberra ACT 0200, Australia*

[2]*Centre for Micro-Photonics, Swinburne University of Technology, Hawthorn, Vic 3122, Australia*

*e-mail: eugene.gamaly@anu.edu.au; sjuodkazis@swin.edu.au; andrei.rode@anu.edu.au*

It was demonstrated during the past decade that ultrashort intense laser pulse tightly focused deep inside a transparent dielectric generates the energy density in excess of several MJ/cm$^3$. Such energy concentration with extremely high heating and quenching rates leads to unusual solid-plasma-solid transformation paths overcoming kinetic barriers to formation of previously unknown high-pressure material phases, which are preserved in the surrounding pristine crystal. These results were obtained with the pulse of Gaussian shape in space and in time. Recently it was shown that the Bessel-shaped pulse could transform much larger amount of a material and allegedly create even higher energy density than that was achieved with the Gaussian (GB) pulses. Here we present a succinct review of previous results and discuss the possible routes for achieving higher energy density employing the Bessel beams (BB) and take advantage of its unique properties.

## 1. Microexplosion studies with Gaussan-shaped beam

It was proven during the last decade that the short intense laser beam with the Gaussian spatial and temporal intensity profile tightly focussed inside a transparent crystal generates the energy density of several MJ/cm$^3$. The pressure produced is in excess of a few TPa exceeding the strength of any existing material (diamond has the highest Young modulus of 1 TPa = 1 MJ/cm$^3$). The laser pulse, 150 fs, 100-200 nJ, 800 nm, tightly focussed inside sapphire with microscope lens (NA = 1.4) creates the solid density plasma at the temperature of a few tens electron Volts ($\sim$ 500,000 K) with the record-high heating rate of $10^{18}$ Kelvin/sec [1,2]. It was found that the novel (previously unobserved) high-pressure phases of Aluminium and Silicon were formed [3,4]. Pressure/temperature conditions created in the micro-explosion are similar to those in hot cores of stars and planets ('primeval soup' or Warm Dense Matter). The material converted to high pressure/temperature solid density plasma is then transformed into the novel solid phase during the ultra-fast cooling and re-structuring. The major difference from the core-star conditions is the record-fast cooling ($\sim 10^{16}$ Kelvin/sec) from plasma state to solid state. In the previous experiments the study of the pressure-affected materials was produced *post mortem*, well after the end of the pulse when transformed material was cooled down to the ambient conditions. The structure of laser-transformed material was determined by the x-rays diffraction from synchrotron [3] and with the electron diffraction [4].

## 1.1. Novelty of the phase transformation path during and after confined microexplosion

Solid transforms to a solid-density plasma state ($T_e$ ~50 eV) during the pulse time shorter than all energy relaxation times. Strong shock wave (SW) starts propagating from the energy deposition region several picoseconds after the pulse due to energy transfer from electrons to the ions. The shock wave decelerates and converts into the sound wave in the surrounding cold pristine crystal. The phenomenon is similar but not identical to the underground nuclear explosion: the massless energy carriers (photons) deliver the energy inside a transparent crystal without changing the atomic and mass content of a material. All laser-affected material is expelled from the energy deposition area by the combined action of shock and rarefaction waves, forming a void surrounded by the shell of material compressed against the surrounding cold pristine crystal. The material returns from the high-pressure plasma state (high entropy, chaotic) to the ambient conditions at room temperature/pressure, however attaining the phase state different from the initial solid state. In all known methods of high pressure phase formation the initial crystalline structure is re-structured, i.e. the atoms are moved from the initial arrangement to the new positions under the action of high pressure. During the transformation path under confined micro-explosion the initial state of a crystal is completely destroyed and forgotten. The irradiated material converted into the chaotic mixture of ions and electrons at high temperature. Therefore relaxation to the ambient conditions occurs along the unknown paths going through the metastable intermediate equilibrium potential minima. The theoretical (computational, modified DFT-studies) during the last decade searched for the possible paths of material transformations under high pressure from the initially chaotic (stochastic) state [5]. These studies uncovered many physically allowed paths for formation of multiple novel phases (including incommensurable phases) from the initially chaotic state. The confined micro-explosion method now is the only practically realised way for formation novel material phases from theplasma state preserving the transformed material confined inside the pristine crystal ready for the further structural studies.

## 1.2. Limitations of the confined micro-explosion method with the Gaussian Beam

There are limitations on the energy density and amount of laser-affected material in confined micro-explosion generated by tightly focused Gauss beam. The main limitation imposed by diffraction: the radius of the diffraction-limited focal spot is [2,6]: $r_{Airy} = 0.61\lambda / NA$. For 800 nm and $NA = 1.4$ one gets $r_{foc} \approx 0.35$ µm and focal area of 0.38 µm$^2$. The absorption length in dense plasma equals to ~30 nm giving the energy deposition volume ~$10^{-14}$ cm$^3$. With absorbed energy around 100 nJ the absorbed energy density amounts to $10^7$ J/cm$^3$ = 10 TPa. The number of laser-affected atoms constitutes around $10^{11}$ atoms (a few picograms) making structural studies complicated.

Therefore, the questions arise: is it possible to increase the absorbed energy density and/or increase the amount of the laser-affected material and thus the amount of the novel phase? Preliminary studies have shown that it is very difficult to overcome the energy density of several MJ/cm$^3$ (several TPa of pressure) and increase the amount of laser-affected material using tightly focused Gauss beam. First, the ionisation wave moving towards the laser pulse with increasing intensity increases the absorbing volume and limits the energy density [12]. Moreover, the experiments with increasing laser pulse energy demonstrated that at the energy per 150 fs pulse of 200 nJ the cracks surrounding the focal area destroyed the regular void formation [2].

Diffraction free Bessel beams (BB) arisen a hope to achieve higher energy density and larger amounts of the material affected. Below we describe the recent progress made with these studies. Then we describe some effects (and unresolved problems), which solutions may lead to further increase of the absorbed energy density.

## 2. Status of the BB-transparent crystal interactions

It was demonstrated recently that the BB (150 fs, 2 μJ) focused inside sapphire produced the cylindrical void of 30 μm length and 300 nm diameter [7]. The void volume, $V= 30\mu m \times \pi r^2 = 2.12 \times 10^{-12}$ cm$^{-3}$, appears to be two orders of magnitude larger than that generated by the GB. The conclusions based solely on the void size measurements and on the energy and mass conservation laws without any *ad hoc* assumptions about the interaction process are the following [8]. The material initially filled the void was expelled and compressed into a shell by the high-pressure shock wave. The work necessary to remove material with the Young modulus $Y$ from volume $V$ equals at least to $Y \times V = 0.848$ μJ ($Y = 4 \times 10^5$ J/cm$^3$ –the Young modulus of sapphire). This is the evidence *of strong ( > 40%) absorption* of the pulse energy. In order to generate a strong shock wave capable expelling such amount of material the absorbed energy should be concentrated in the central spike with the much smaller diameter than that of the void (still unknown either theoretically or experimentally).

The unique features of the diffraction-free Bessel beam spatial distribution of intensity in the focal area allow understanding some experimental finding and indicate to new problems and opportunities. The spatial distribution of intensity across the cylindrical focal volume in a transparent medium unaffected by light and observed experimentally is close to the Durnin's solution, $J_0^2\left(k_r r\right)$ [8]: the central spike surrounded by circular bands with the maximum of intensity on the axis approximately 5 times higher than in the next band. The length of the focal area is defined by the parameters of device creating the BB (axicon, SLM or others).

Under the action of intense pulse the ionization breakdown occurs early in the pulse time near the central spike where intensity is a maximum. The studies of interaction process of intense BB at intensity above the ionization threshold are absent for the best of our knowledge. Estimates, suggestions and problems relating to the formation of the intensity distribution and interaction process based on the studies of confined micro-explosion and intense short pulse interactions with dielectrics are presented below.

The ionisation threshold occurs at zero-real-permittivity surface, separating plasma in a region close to the axis and dielectric areas outside. The spatial distribution of the excited permittivity changes from positive to negative values. The light interacts with electrically inhomogeneous medium after breakdown while the ions remain at unperturbed positions. It is known that when the electric field direction in the excitation beam coincides with the permittivity gradient in the target the interaction has resonance character and electric field has a maximum near the resonance point [see Ref.8 and the references therein]. In this case the energy flux directed along the permittivity gradient (in radial direction to the axis) is created.

The ideal diffraction free beam is the monochromatic Bessel beam [9], created via superposition of plane waves whose wave vectors are evenly distributed over the surface of a cone. It was shown [11] that the BB can be presented as the result of the interference of two conical Hankel beams [11], carrying equal amounts of energy towards and outwards the beam axis, and yielding no net transversal energy flux in the BB. Interference of two Hankel beams with different amplitudes creates unbalanced BB where the net radial energy flux appears. Unbalancing creates the inward radial energy flux from the conical tails of the beam. The study of stability in the frame of NLSE equation revealed that the Bessel-like solutions in pure Kerr media are unstable [10].

It was shown [10] that in a medium with the weak Kerr nonlinearity and nonlinear energy losses the inward radial energy flux is created to refill the nonlinearly absorbed energy near the central axis keeping the BB balanced and stationary.

In the interaction of intense short pulse BB with transparent dielectric at the intensity below the ionization threshold the BB apparently retains its balanced structure. However after the plasma formation in the central core the strong resonance absorption occurs at the dielectric/plasma (zero real part of the permittivity) boundary. The electric field has a maximum there, which is higher than in simple plasma due to large electrostatic field [8]. The back reflection from the real-permittivity-zero surface changes the interference pattern responsible for formation diffraction-free BB. Therefore the BB in dielectric before the zero-permittivity surface allegedly becomes unbalanced and the inward energy flux is created. One may conjecture that the BB becomes unstable tending to focus onto the cylindrical axis thus creating the energy density higher than tightly focused but diffraction limited Gaussian beam.

Experiments demonstrated that short intense BB could affect much larger amount of material producing solid-plasma-solid transformation (direct measurements) at allegedly pressure of several TPa (conclusions on the basis of the analysis of the experiments) [7,8].

The Chinese group measured the average speed of the shock wave, $v_{sw} \approx 60$ km/s, during the cylindrical micro-explosion, generated by the BB in sapphire, by the pump-probe technique (S. Juodkazis private communication from the conference presentation). The estimate of the driving pressure based on this measurements, $P_{sw} \cong \rho_0 v_{sw}^2 = 14.4$ TPa ($\rho_0$ is the initial mass density of sapphire), gives the direct experimental evidence of the extreme energy density created by the BB in the focal volume.

There are indications from the theoretical studies [8] that the originally stable diffraction free BB at high intensity in the presence of strong ionization nonlinearity may become unstable. Now it is difficult to conclude if it may happen in a way similar to the self-focusing instability with Kerr-like non-linearity (rather not because the paraxial approximation is invalid in this case) or similar to the instability of two unbalanced Hankel beams, which seems more relevant to the case (again the ionization non-linearity should be accounted for).

The oblique incidence, inherent for the formation of the BB and long focus, implies the possibility of the surface wave (plasmon) formation and propagation along the zero-real-permittivity surface at the same time with the plasmon moving radially due to the resonance absorption. The plasma wave may converge to the axis contributing to increase in the absorbed energy density. One may conjecture if it might be relevant to some kind of the Langmuir collapse.

It would be crucially important to find electric field distribution up to the central axis in order to determine the absorbed energy density. It means solution of the Maxwell equations in cylindrical geometry coupled to material equations accounting for the change in the permittivity (electrons' number density and collision rate) in accord to intensity in any space/time point. It is a formidable task however it can be clearly formulated for the numerical solution. Different approximations may also be discussed.

Theoretical modelling of the cylindrical explosion after the energy deposition of the BB beam inside a narrow on-axis cylinder also can be performed in the frame two-temperature plasma hydrodynamics in cylindrical geometry in a way similar to that was done with the Gauss beam in spherical geometry [2].

## 3. Conclusions

In conclusion we should state that the further progress in achieving and steering the high energy density strongly depends on the future pump-probe experiments, which will register with time/space resolution the history of the BB generated micro-explosion, processes of returning to the ambient state and new phases formation. It is worth to show the time and space scales for the succession of events comprising such a history that might in some approximation be extracted from the previous studies [2, 4 (Suppl. material), 8].

Let's suggest the BB, 2 μJ, 800nm, 150 fs, impinges sapphire crystal several tens of microns thick creating a focal region of ~ 30 μm long at the ten microns depth from the outer surface of a sample. The stages of successive transformations are the following, time count starts at the beginning of the pump pulse.

1. Low intensity stage before ionisation threshold lasts a few fs as the beginning of the pulse.

2. As the ionisation threshold is attained, the cylindrical plasma region is created at the axis of the focal region with the diameter less than a micron. One should note that the full length of the focal region of 30 μm is reached at the end of the pulse, assuming that light propagates as in unaffected sapphire with the speed of $c/n$ ~$2 \times 10^{10}$ cm/s).

3. The diameter of the energy absorption region to the end of the pulse allegedly might be around the doubled absorption length in dense plasma ~ 60 nm.

4. The shock wave is created after the energy transfer from electrons to ions in 7-10 ps time span.

5. The shock wave propagates during 4-6 ps until it is converted into the acoustic wave, effectively stopped by the cold pressure of the crystal (~ Young modulus of sapphire). The void surrounded by the shell of compressed material is formed by the rarefaction wave.

6. Thermal wave of conventional heat conduction spreads into the laser-unaffected crystal cooling the laser-affected area down to the ambient conditions during tens of nanoseconds. The material re-structuring occurs most probably during the stages 5 and 6. The whole area affected by the heat from the laser-heated region is a cylinder with length around 32-34 microns with a diameter about 2-4 microns.

Thus, the whole area affected by the shock and heat waves from the energy deposition region is a cylinder 30 micron long and a few microns in diameter. Time span for the whole processes of material transformation is around tens of nanoseconds. Recent arrival of X-ray free electron lasers with pulse duration of $15 - 25$ fs and the photon energy ~8 KeV currently available at EuroXFEL at DESY in Hamburg, and with up to 17 KeV to be expected in near future at the SLAC National Acceleration Laboratory in Stanford, as well as SACLA at Spring-8 at Riken Institute in Japan are the ideal sources to be used as a probe pulses to uncover the process of formation such unusual material states.

To conclude the light (or x-ray) probe with sub-picosecond duration and sub-micron spatial resolution may shed light on the unusual formation of novel high-pressure phases starting from "primeval soup" (Warm Dense Matter) to the solid state at the ambient being preserved on the laboratory table top and confined inside a bulk of pristine crystal ready for the further structural studies.

## Acknowledgement

We acknowledge funding from the Australian Government through the Australian Research Council's Discovery Projects funding scheme (DP170100131).